\documentstyle[aps,twocolumn,prl,amsfonts]{revtex}
\begin{document}
%
%
%%%%%%%%%%%%%%%%%%%  Title and Abstract  %%%%%%%%%%%%%%%%%%%%%%%%%%%%%%%%%%
%
%
\draft
\title{Dynamics by White--Noise Hamiltonians}
\author{Werner Fischer, Hajo Leschke, and Peter M\"uller}
\address{Institut f\"ur Theoretische Physik,
Universit\"at Erlangen--N\"urnberg,\\
Staudtstra{\ss}e 7, D--91058 Erlangen, Germany}
\date{13 April 1994}
\maketitle
\begin{abstract}
{A new class of random quantum--dynamical systems in continuous space is
introduced and studied in some detail. Each member of the class is
characterized by a Hamiltonian which is the sum of two parts.
While one part is deterministic, time--independent, and quadratic, the
Weyl--Wigner symbol of the other part is
a homogeneous Gaussian random field
which is delta correlated in time and arbitrary, but smooth in
position and momentum. Exact expressions for the time evolution of both
(mixed) states and observables averaged over randomness are obtained.
The differences between the quantum and the classical
behavior are clearly exhibited.
As a special case it is shown that, if the deterministic part corresponds
to a particle subjected to a constant magnetic field,
the spatial variance of the averaged state grows diffusively
for long times independent of the initial state.}
\end{abstract}
\pacs{05.40.+j, 05.60.+w, 72.70.+m}
\narrowtext
%
%
%%%%%%%%%%%%%%%%%%%%%%%%%%   Main part  %%%%%%%%%%%%%%%%%%%%%%%%%%%%%%%
%
%

The spatial spreading of a state under the free time evolution is a
well--known and fundamental phenomenon in non--relativistic quantum and
classical mechanics \cite{Farina}. In order to make a quantitative
statement, let $ \sigma _{t}^{2} $ denote the variance of the position
at time $ t $ of a spinless point particle moving in continuous space.
It is assumed that the particle was prepared initially in
some state, which is normalized but not necessarily pure.
For a free particle a simple calculation then
shows that $ \sigma _{t}^{2} $ increases asymptotically for large $ t $
as $ \sigma _{t}^{2} \sim t^{2}\varrho_{0}^{2}/m^{2} $. Here
$ m>0 $ is the mass of the particle and $ \varrho_{0} ^{2} $ is the variance
of its momentum in the initial state. This relation holds both in the
quantum and in the classical case. However, classical states,
e.g.\ pure ones, may have a sharp momentum, that is,
$ \varrho_{0}^{2}=0 $, whereas $ \varrho_{0}^{2} >0 $ for
all quantum states, including the pure ones (``wave--packet spreading'').
It is clear that $ \sigma _{t}^{2} $ can also be calculated
exactly for a
time evolution governed by a more general Hamiltonian being at most
quadratic in momentum and position \cite{Styler}.

A variety of physical systems, whose properties are unknown in
detail, are successfully modeled by a Hamiltonian with a random part
added to a (simple) deterministic part. Thus it is a challenging problem
to study the properties of the random variable $ \sigma _{t}^{2} $
and of related key quantities in these models. For recent
investigations devoted to the question, how the ballistic long--time
behavior of $ \sigma _{t}^{2} $ of the free particle is modified by
adding a Gaussian time--dependent random potential, see
\cite{JayannavarK,GolubovicF,Jayannavar} and
references therein.
A stepping stone in this direction was the work of
Jayannavar and Kumar \cite{JayannavarK}, who---in building on treatments
of lattice models \cite{OvchinnikovE}---exploited the
simplifying feature of a
vanishing correlation time (``Gaussian white--noise potential'').
Their main result concerns the quantum case with a particular pure
initial state.
They derived an exact
expression for the spatial variance $ \Sigma _{t}^{2} $ of the
averaged state at time $ t $ and found $ \Sigma _{t}^{2} \sim t^{3} $
for large $ t $.
(Note that the averaged spatial variance $ \overline{\sigma _{t}^{2}} $
never exceeds $ \Sigma _{t}^{2} $.)
Interesting attempts to incorporate a non--zero correlation
time by the use of perturbative methods---with partially conflicting
results---can be found in
\cite{GolubovicF}.

However, several problems of considerable interest have not been tackled
so far. Firstly, in order to describe the effects of
externally applied force fields, one must not restrict the deterministic
part of the Hamiltonian to that of a free particle. For example, we
will see that the presence of a constant magnetic field leads to a
diffusive behavior in the sense that $ \Sigma _{t}^{2} \sim t $,
a result with some relevance for magneto--transport theory. Secondly, the
random part can be generalized to cover the case of a
momentum--dependent (in other words, non--local) random potential,
which is the continuous--system analog of off--diagonal disorder in lattice
systems \cite{OvchinnikovE}. This
is of interest, for example, to caricature the effective motion of a
test particle due to inelastic scattering by
the irregular motion of other particles.
It suggests itself also from the point of view of Hamiltonian mechanics.
And lastly, we will show that the above--mentioned noise--induced
results are neither affected by quantum fluctuations
nor do they depend on the initial state.

In fact, it is the main purpose of the present Letter to demonstrate
that there is a rather general class of Gaussian white--noise
Hamiltonians for which one can obtain exact and explicit results on the
averaged time evolution. Yet before we describe this class in
detail, it seems adequate to comment on the representation we are going to
use throughout.

We consider a quantum--mechanical system which, for
simplicity, has the Euclidean line $ {\Bbb{R}} $ as its configuration
space. The extension to the $ d $--dimensional Euclidean space
$ {\Bbb{R}}^{d} $ is merely a matter of notation.
Since the random part of the Hamiltonian will be allowed to depend on
both position and momentum, it is convenient to characterize its
properties in terms of those of an associated random
function on classical phase space $ {\Bbb{R}} \times {\Bbb{R}} $.
Therefore, and in order to treat the classical limit with low effort,
it is only consequent to represent the quantum
system entirely in phase space. The representation we
choose is the one dating back to ideas of
Weyl, Wigner, and Moyal \cite{WeylWigner},
where a quantum operator $ \hat{f} $ acting in the Hilbert space of
square--integrable
functions on $ {\Bbb{R}} $ is represented uniquely by its
symbol, that is, by the phase--space function
\begin{equation} \label{symbol}
f(p,q) := \int_{{\Bbb{R}}} dr\;
e^{ipr/\hbar}\;
\langle q-r/2|\,\hat{f}\,|q+r/2 \rangle\;.
\end{equation}
We recall that the symbol of the standardized commutator
$ (i/\hbar)(\hat{f}\hat{g}-\hat{g}\hat{f}) $ of two operators
$ \hat{f} $ and $ \hat{g} $ is the Moyal bracket
of their symbols $ f $ and $ g $
$$
[ f,g ](p,q) :=  f(p,q)\; \frac{2}{\hbar}\;
\sin\left\{ \frac{\hbar}{2} \left(
\stackrel{\leftarrow}{\partial_{p}}
\stackrel{\rightarrow}{\partial_{q}} -
\stackrel{\leftarrow}{\partial_{q}}
\stackrel{\rightarrow}{\partial_{p}} \right)
\right\} \,g(p,q) \;.
$$
For non--polynomial $ f $ and $ g $ it is often advantageous to rewrite
$ [f,g] $ as a Fourier--integral expression.
Of course, in the classical limit, when Planck's constant $ \hbar $
tends to zero, the Moyal bracket reduces to the Poisson bracket.
An observable, corresponding to a self--adjoint operator,
has a real symbol $ a $.
A quantum state is represented by a Wigner density $ w $
which, by definition, is $ (2\pi \hbar)^{-1} $ times the symbol
of the corresponding ``density matrix'', that is, of a positive
unit--trace operator.
The expectation value of the observable $ a $ in the state $ w $ is then
given by the scalar product
$$
\langle w,a\rangle := \int_{{\Bbb{R}} \times {\Bbb{R}}}
dp\,dq\; w(p,q)a(p,q)\;.
$$
We recall that $ |w(p,q)| \le (\pi \hbar)^{-1} $ and that $ \langle
w,1\rangle =1 $.
Moreover, $ \langle w,w\rangle \le (2\pi \hbar)^{-1} $ with equality
if and only if $ w $ represents a pure quantum state.
In the classical limit a quantum state $ w $ converges (weakly) to a
probability density on phase space, that is, to a classical state.

Now we are in a position to introduce a dynamics governed by a
Hamiltonian on phase space which we call white--noise Hamiltonian
\begin{equation} \label{model}
H(p,q) + N(p,q;t) \; .
\end{equation}
By definition, the Hamiltonian operator is obtained from (\ref{model}) by
inverting (\ref{symbol}), that is, by Weyl ordering.
For sim\-pli\-ci\-ty, the deterministic part $ H(p,q) $ is supposed to be
time--independent and at most quadratic in $ p $ and $ q $.
The random part $ N(p,q;t) $ is supposed to be a
Gaussian white--noise field with mean zero and covariance
\begin{equation}\label{cov}
\overline{N(p,q;t) N(p',q';t')}= C(p-p',q-q')\delta (t-t') \;.
\end{equation}
Here the overbar denotes averaging with respect to the probability
distribution of $ N $ and homogeneity is assumed just for brevity.
By its probabilistic origin, the covariance function $ C $ may be any even
phase--space function
with a non--negative (symplectic) Fourier transform
$$
\widetilde{C} (x,k) := \int_{{\Bbb{R}} \times {\Bbb{R}}}
\frac{dp\,dq}{(2\pi)^{2}}\; C(p,q)
\cos (x p - k q) \; \ge 0\;.
$$
For later purpose we will assume that $ C $ is suffi\-cient\-ly smooth,
equivalently, that the probability density
$ \widetilde{C}(x,k)/C(0,0) $ has moments of sufficiently high order.

In the Schr\"odinger picture the time evolution $ w_{0}\mapsto w_{t} $
of a given initial state $ w_{0} $ is determined by the
stochastic quantum Liouville equation \cite{Kubo} associated with
(\ref{model})
\begin{equation} \label{lvn}
\partial_{t} w_{t}
= [ w_{t}, H ] + [ w_{t}, N(t) ] \;.
\end{equation}
Here the bracket $ [ w_{t}, H ] $ is in fact a Poisson bracket
due to the quadratic nature of $ H $.
In order to derive an equation of motion for the averaged
state $ \overline{w_{t}}(p,q):=\overline{w_{t}(p,q)} $ from (\ref{lvn}),
we follow essentially
the earlier treatments in \cite{OvchinnikovE,JayannavarK} and
perform a functional integration by parts
with respect to the Gaussian average
\cite{Novikov}. In doing so, we think of the Dirac delta function
in (\ref{cov}) as being approximated by a sequence of smooth covariance
functions with correlation time tending to zero, which amounts to the
Stratonovich interpretation \cite{Mortensen} of (\ref{lvn}).
The final result can be
cast into the form of a
linear integro--differential equation
\begin{eqnarray} \label{ide}
\lefteqn{\partial_{t} \overline{w_{t}}(p,q)=
[ \overline{w_{t}},  H](p,q)}\nonumber\\
& & +
\frac{1}{\hbar^{2}} \int_{{\Bbb{R}} \times {\Bbb{R}}} dx\, dk\;
\widetilde{C}(x,k)
\left\{\overline{w_{t}}(p+\hbar k,q+\hbar x) -
\overline{w_{t}}(p,q)  \right\}\nonumber\\
\end{eqnarray}
which is valid for $ t>0 $ and has to be supplemented by the initial
condition $ \overline{w_{0}} = w_{0} $. Several remarks are in
order:

$\bullet$ Eq.\ (\ref{ide}) is a substantial generalization of
the main result of
\cite{JayannavarK}. In the special case of a free deterministic part,
$ H = p^{2}/2m $, and a momentum--inde\-pen\-dent random potential,
$ \widetilde{C}(x,k) = \delta (x) \gamma (k) $, it reduces to an equation
which is equivalent to Eq.\ (8) in \cite{JayannavarK}.
Furthermore, the subsequent treatment of Eq.\ (8) in \cite{JayannavarK}
is restricted to a pure initial state represented by a joint Gaussian
$ w_{0} $ with
$ \langle w_{0},p\rangle = \langle w_{0},q\rangle =
\langle w_{0},pq\rangle = 0,\;
\langle w_{0},q^{2}\rangle = \sigma_{0}^{2} \not= 0 $, and
$ \langle w_{0},p^{2}\rangle = (\hbar/2\sigma _{0})^{2} $.

$\bullet$ The averaged time evolution $ {\cal T} _{t}:
w_{0}\mapsto\overline{w_{t}} $ given by (\ref{ide}) provides an example
of a quantum--dynamical semigroup \cite{Spohn,Alicki} which is
monotone mixing--increasing.
By this we mean that $ {\cal T} _{t} $ maps Wigner densities linearly to
Wigner densities,
$ {\cal T} _{t}\circ{\cal T} _{t'} = {\cal T} _{t+t'} $ for all
$ t,t'\ge 0 $, $ {\cal T} _{0}=~identity $, and
$ \partial_{t}\,
\langle {\cal T}_{t} (w_{0}), {\cal T} _{t} (w_{0}) \rangle \le 0 $
(with equality only in the uninteresting case where the
covariance function $ C $ is a constant).
The inequality follows from scalar multiplication of (\ref{ide})
by $ \overline{w_{t}}={\cal T}_{t} (w_{0}) $, by observing
$ \langle \overline{w_{t}},[\overline{w_{t}},H]\rangle = 0 $, the
Cauchy--Schwarz inequality and $ \widetilde{C}\ge 0 $.
To summarize, the average over randomness has turned the fully
reversible quantum Liou\-ville equation (\ref{lvn}) into the equation
(\ref{ide}) with coherence--destructing irreversible behavior.

$\bullet$ Interestingly enough, modified quantum--dynami\-cal equations
similar to (\ref{ide}) are discussed in very different branches
of physics. These include quantum theories not only of certain disordered
systems, but also of the measurement
process \cite{GhirardiR}, of Markovian transport \cite{Spohn}, and of the
evaporation of black holes \cite{BanksS}.

$\bullet$ As for the generator of the semigroup, one deduces from Eq.\
(\ref{ide}) that
\begin{equation}\label{semi}
{\cal T} _{t}= e^{ -t({\cal L} + {\cal N})}
\end{equation}
with the unperturbed Liouville operator
\begin{equation}\label{L0}
{\cal L}:= (\partial_{p}H)\partial_{q} -
(\partial_{q}H)\partial_{p}
\end{equation}
and the noise--induced, irreversibility causing operator
$$
{\cal N}:= \hbar^{-2}\{ C(0,0) - C(-i\hbar\partial_{q},
i\hbar\partial_{p})\}\;.
$$

$\bullet$ For the derivation of explicit results it is often useful to
isolate the unperturbed time evolution in (\ref{semi}) according to
standard perturbation theory
\begin{equation} \label{dirac}
{\cal T} _{t}=
e^{-t{\cal L}}\exp\left\{ -\int_{0}^{t}ds\; e^{s{\cal L}}\;
{\cal N}\;e^{-s{\cal L}}\right\}\;.
\end{equation}
We note that due to the  quadratic nature of $ H $ the operators
$ e^{s{\cal L}}\;{\cal N}\;e^{-s{\cal L}} $ commute at
different times $ s $ and can be written more explicitly as
\begin{equation}\label{differ}
e^{s{\cal L}}\;{\cal N}\;e^{-s{\cal L}} =
\hbar^{-2}\{ C(0,0) - C(-i\hbar {\cal K}_{s},
i\hbar {\cal X}_{s})\}\;.
\end{equation}
Here we have introduced the time--dependent
first--order differential operators
$ {\cal K}_{s}:= (\partial_{q} e^{-s{\cal L}}q)\partial_{q}
+ (\partial_{q} e^{-s{\cal L}}p)\partial_{p} $ and
$ {\cal X}_{s}:= (\partial_{p} e^{-s{\cal L}}p)\partial_{p}
+ (\partial_{p} e^{-s{\cal L}}q)\partial_{q} $,
whose coefficients are obtained from the phase--space
trajectory of the unperturbed problem as indicated,
and do not depend on $ p $ and $ q $.

$\bullet$ By using (\ref{dirac}) and (\ref{differ})
it is straightforward to derive a
Fourier--integral expression for the integral kernel
$ {\sf T}_{t}(p,q|p',q') $ of $ {\cal T}_{t} $, which is the solution
of (\ref{ide}) with initial condition
$ {\sf T}_{0}(p,q|p',q') = \delta (p-p') \delta (q-q') $.
Since we will not need this expression below, we omit it.

$\bullet$ The quantum--dynamical semigroup $ {\cal T}_{t} $
admits also a purely classical interpretation.
This is because it is positivity
preserving, which follows from (\ref{dirac}) and the fact that
$ C(-i\hbar\partial_{q},i\hbar\partial_{p}) $ and
$ {\rm e}^{\pm s{\cal L}} $ are positivity preserving.
Therefore, $ {\cal T}_{t} $ maps classical states to classical states.
In other words, its integral kernel
$ {\sf T}_{t}(p,q|p',q') $ can be interpreted as the
transition density of a stationary, in general non--continuous,
Markov process in phase space and
Eq.~(\ref{ide}) may be viewed as the associated classical kinetic equation.
In fact, up to the drift arising from $ H $, it is a linear
Boltzmann equation with a (homogeneous) stochastic kernel \cite{LasotaM}.
It is used, e.g., in quasi--classical theories of charge transport in
semiconductors \cite{MarkowichR}.

$ \bullet$ Assuming an $ \hbar $--independent covariance function
$ C $, one has the
expansion
\begin{equation}\label{L1}
{\cal N} = - D_{0,2} \partial^{2}_{p} - D_{1,1}\partial_{p}
\partial_{q} - D_{2,0} \partial^{2}_{q} +
{\cal O}(\hbar^{2}\partial^{4})\;,
\end{equation}
where the three constants $ D_{0,2} $, $ D_{1,1} $ and $ D_{2,0} $ as
defined through
$ D_{\mu ,\nu }:= (-i\partial_{p})^{\mu }
(i\partial_{q})^{\nu } C(0,0) / \mu !\, \nu ! $
reflect the curvature of $ C $ at the origin and
obey the inequalities $ D_{0,2}\ge 0 $ and $ 4 D_{2,0} D_{0,2}\ge
D_{1,1}^{2} $ due to $ \widetilde{C}\ge 0 $.
As a consequence, in the classical limit Eq.\ (\ref{ide}) reduces to a
Fokker--Planck--type equation in phase space with drift and
diffusion as given by (\ref{L0}) and (\ref{L1}).

Now we return to the problem posed in the beginning of this Letter,
namely to evaluate the averaged expectation value
$ \overline{\langle w_{t},a\rangle} $ of a simple observable $ a $ at
time $ t $, given the initial state $ w_{0} $.
For this purpose it is useful
to switch to the Heisenberg picture according to
\begin{equation}\label{heisenberg}
\overline{\langle w_{t},a\rangle}=\langle\overline{w}_{t},a\rangle=
\langle{\cal T}_{t} (w_{0}),a\rangle =:\langle w_{0},
{\cal T} _{t}^{*} (a)\rangle\;.
\end{equation}
The thus defined adjoint semigroup $ {\cal T} _{t}^{*} $ can be obtained
from (\ref{semi}) or (\ref{dirac}) by reversing the sign of $ {\cal L} $.

To be more specific, we first choose the
de\-ter\-mini\-stic part of the Hamiltonian (\ref{model})
to be that of a free particle,
$ H= p^{2}/2m $. Taking the observables
$ p $, $ q $, $ p^{2} $, $ pq $, $ q^{2}$, and $ q^{4} $
as examples, one then finds explicitly
\begin{mathletters}\label{frei}
\begin{eqnarray}
{\cal T}_{t}^{*} (p)^{~} & = & p \; , \quad
{\cal T}_{t}^{*} (q) = q + tp/m , \\[1.5mm]
\label{p2}
{\cal T}_{t}^{*} (p^{2})  & = & p^{2} + 2tD_{0,2}  \;, \\[1.5mm]
{\cal T}_{t}^{*} (pq)  & = & p(q+tp/m) + tD_{1,1}
+ t^{2}D_{0,2}/m\;, \\[1.5mm]
\label{q2}
{\cal T}_{t}^{*} (q^{2}) & = & (q + tp/m)^{2} + 2\, Q_{3}(t) \;, \\[1.5mm]
{\cal T}_{t}^{*} (q^{4}) & = & (q+tp/m)^{4}
+12\{ (q+tp/m)^{2} Q_{3}(t)\nonumber\\
& &\hspace*{2.3cm} + 2\hbar^{2}Q_{5}(t) + (Q_{3}(t))^{2}\}.
\end{eqnarray}
Here $ Q_{\mu}(t) :=  \sum_{\nu =1}^{\mu } t^{\nu }
D_{\mu -\nu ,\nu -1}\,m^{1-\nu}/\nu $
is a polynomial of (maximum) degree $ \mu $ in time.
Quantities such as the spatial variance of the averaged state
$ \Sigma_{t}^{2}:=\langle w_{0}, {\cal T}_{t}^{*} (q^{2}) \rangle  -
\langle w_{0}, {\cal T}_{t}^{*}(q)\rangle ^{2} $ or the
averaged mean--square
displacement $ \Delta_{t}^{2}:= \langle w_{0}, {\cal T}_{t}^{*} (q^{2})
-2q {\cal T}_{t}^{*} (q) + q^{2} \rangle $ at time $ t $,
may now be obtained immediately.\end{mathletters}

The exact results (\ref{frei}) illustrate important features valid for
general quadratic $ H $: Observables which are linear
in $ p $ and $ q $ are not affected by the white noise
$ N $. Moreover, noise--induced terms in averaged expectation values
(\ref{heisenberg}) of quadratic observables are independent of the
initial state. Assuming an $ \hbar $--independent covariance function
$ C $, noise--induced effects affected by quantum fluctuations occur
only for observables of at least fourth order in $ p $ and
$ q $. However, very special situations are needed for quantum effects
to show up in the leading term for long times.
For example, taking the
observables $ p^{n} $ or $ q^{n},\; 4\le n {\rm ~integer} $, one must
require the phase--space trajectories of $ H $ to grow exponentially in
time.
This remark contradicts certain expressions in
\cite{Jayannavar}, since the underlying ``correlation functions''
considered there do not have a positive Fourier transform, and are
therefore physically insignificant.

White--noise Hamiltonians may reveal a diffusive behavior in a weak
sense, that is, $ {\cal T}_{t}^{*} (p^{2}) \sim t $ and/or
$ {\cal T}_{t}^{*} (q^{2}) \sim t $ for long times.
The simplest case for weak diffusion to occur
in both momentum and position corresponds to $ H=0 $, as follows
from (\ref{p2}) and (\ref{q2}) in the limit $ m \to \infty $.
It occurs also if the deterministic part describes a harmonic
oscillator and, more strikingly, in the case of a particle with
electric charge
$ -e $ moving in the Euclidean plane $ {\Bbb{R}}^{2} $ under the
influence of a perpendicular constant magnetic field of strength
$ |m\omega /e| $. In the latter case we choose
$ H=(2m)^{-1}\left\{ (p_{1}-m\omega q_{2}/2)^{2} +
(p_{2}+m\omega q_{1}/2)^{2}\right\} $ and, for the sake of brevity, we
assume the covariance function $ C({\bf p},{\bf q}) $ to depend only on
the absolute values $ |{\bf p}| $ and $ |{\bf q}| $. By a simple
extension of the presented methods to higher dimensions,
one then finds for the averaged squared position at time $ t $
\begin{eqnarray} \label{bfeld}
{\cal T}_{t}^{*} ({\bf q}^{2}) & = &
\left({\rm e}^{t{\cal L}}{\bf q}\right)^{2} -t\,\bigg\{
\left( 1 + \frac{\sin \omega t}{\omega t}\right)
\partial^{2}_{p_{1}}C({\bf 0},{\bf 0})
\nonumber\\
& &  + \left( 1 -  \frac{\sin \omega t}{\omega t}\right)
\left(\frac{2}{m\omega }\right)^{2}
\partial^{2}_{q_{1}}C({\bf 0},{\bf 0})
\bigg\}\;.
\end{eqnarray}
Here $ {\rm e}^{t{\cal L}}{\bf q} $ is nothing but a
cyclotron orbit associated with $ H $.
Like in lattice models \cite{OvchinnikovE}, but unlike the
free--particle case (\ref{q2}), the leading term
in (\ref{bfeld}) for large $ t \gg \omega^{-1}  $ is influenced
by the noise in both momentum and position.
Note also that the long--time behavior of
$ {\cal T}_{t}^{*} ({\bf q}^{2}) $,
and hence of $ \Sigma _{t}^{2} $ and $ \Delta_{t}^{2} $,
changes suddenly from a cubic to a linear behavior
when turning on a magnetic field.
When the fluctuating environment is modeled by a random
part with a non--zero correlation time, we expect the same trend.
However, the growth should be slower and, as a new feature, it should
depend on the initial state. To verify these conjectures,
non--perturbative methods are needed for controlling the long--time
behavior of quantities as $ \Sigma _{t}^{2} $ and $ \Delta_{t}^{2} $ in
these non--Markovian models.

Not unexpectedly, except for very particular deterministic parts such as
$ H= (p^{2} - Dq^{2})/2m $ with $ D\ge D_{0,2}/D_{2,0}$,
white--noise perturbations lead to a linear increase in time
of the averaged energy
$$
{\cal T}_{t}^{*} (H)  =  H + t\left( D_{0,2}\partial_{p}^{2}H +
D_{1,1}\partial_{p}\partial_{q}H + D_{2,0}\partial_{q}^{2}H \right).
$$
To compensate this effect and, if possible, to allow for an eventual
approach to a stationary (equilibrium) state,
dissipation has to be incorporated, typically by
coupling the white--noise system to a heat bath in the spirit
of~\cite{FordL}. In the context of noise and dissipation, Accardi's
program ``Quantum Stochastic Mechanics'' \cite{Accardi} is very
promising for a deeper understanding of quantum time evolutions.

W.F.\ acknowledges support by the
Evangelisches Studienwerk Villigst (Schwer\-te, Germany)
and P.M.\ acknowledges support by the Studienstiftung des
deutschen Volkes (Bonn, Germany).
%
%
%%%%%%%%%%%%%%%%%%%%%%    References    %%%%%%%%%%%%%%%%%%%%%%%%%%%%%%%%%%%
%
%

%
%
\end{document}